\documentclass[prd,tightenlines,nofootinbib]{revtex4}

\usepackage{epsfig}
\usepackage{mathrsfs}
\usepackage{amsfonts}
\usepackage{color}
\usepackage{amssymb}
\usepackage{citesort}
\usepackage{siunitx}
\usepackage{amsfonts}
\usepackage{amsmath}
\usepackage{graphicx}
\usepackage{amssymb}

\usepackage{dsfont}
\usepackage{bbm}
\usepackage{IEEEtrantools}
\usepackage{hyperref}
\usepackage{tikz}

\newcommand{\leftrightarrowbottom}{
	\begin{tikzpicture}
	\draw[<->] (0,0) -- (2.ex,0);
	\end{tikzpicture}
}
\newcommand{\rightarrowbottom}{
	\begin{tikzpicture}
	\draw[->] (0,0) -- (2.ex,0);
	\end{tikzpicture}
}
\newcommand{\leftarrowbottom}{
	\begin{tikzpicture}
	\draw[<-] (0,0) -- (2.ex,0);
	\end{tikzpicture}
}
\newcommand{\leftrightnablasquare}{\overset{\text{\leftrightarrowbottom}}{\nabla}\hspace{-2pt}\phantom{}^2}
\newcommand{\leftrightnabla}{\overset{\text{\leftrightarrowbottom}}{\nabla}\hspace{-2pt}}
\newcommand{\leftnabla}{\overset{\text{\leftarrowbottom}}{\nabla}\hspace{-2pt}}
\newcommand{\rightnabla}{\overset{\text{\rightarrowbottom}}{\nabla}\hspace{-2pt}}

\allowdisplaybreaks
\newcommand{\eq}{\begin{eqnarray}}
\newcommand{\en}{\end{eqnarray}}

\begin{document}

\title{Finite-volume energy shift of the three-pion ground state}

\author{Fabian M\"uller}
\affiliation{Helmholtz-Institut f\"ur Strahlen- und Kernphysik (Theorie) and Bethe Center for Theoretical Physics, Universit\"at Bonn, 53115 Bonn, Germany}
\author{Akaki Rusetsky}
\affiliation{Helmholtz-Institut f\"ur Strahlen- und Kernphysik (Theorie) and Bethe Center for Theoretical Physics, Universit\"at Bonn, 53115 Bonn, Germany}
\affiliation{Tbilisi State  University,  0186 Tbilisi, Georgia}
\author{Tiansu Yu}
\affiliation{Helmholtz-Institut f\"ur Strahlen- und Kernphysik (Theorie) and Bethe Center for Theoretical Physics, Universit\"at Bonn, 53115 Bonn, Germany}

\begin{abstract}
  Using the framework of nonrelativistic effective field theory, the finite-volume
  ground-state energy shift is calculated up to and including $O(L^{-6})$
  for the system of three pions in the channel with the total isospin $I=1$.
    The relativistic corrections are included
    perturbatively, up to the same order in the inverse of the box size $L$.
    The obtained explicit expression, together with the known result for the system with maximal isospin $I=3$,
  can be used for the extraction of two independent effective three-body couplings
  from the measured ground-state spectrum of three pions.

\end{abstract}

\maketitle

\section{Introduction}	

The extraction of various hadronic observables in the three particle sector is certainly a
challenging task for lattice QCD. Such an extraction has become truly realistic only recently,
owing to the rapid growth of computational resources available, as well as the progress
on the methods and algorithms~\cite{Beane:2007es,Horz:2019rrn,Culver:2019vvu,Fischer:2020jzp,Blanton:2019vdk,Hansen:2020otl,Alexandru:2020xqf}
(see also the closely related
papers~\cite{Romero-Lopez:2018rcb,Romero-Lopez:2020rdq}, where
the simulations were performed within the scalar $\varphi^4$-theory). There is, however,
more to this task than merely carrying out larger simulations which require more
resources. Namely, the extraction of the hadronic parameters, defined in the infinite
volume, from the observables, measured on a finite lattices, has proven to be significantly
more challenging than in the one- and two-particle sectors. The problem has been,
in principle, solved by the three-particle quantization condition, which has been derived
only recently in three different
settings~\cite{Hansen:2014eka,Hansen:2015zga,Polejaeva:2012ut,Hammer:2017uqm,Hammer:2017kms,Mai:2017bge}. In its turn, this development has sparked activities in the field, as seen, e.g., from~\cite{Hansen:2015zta,Hansen:2016fzj,Briceno:2017tce,Sharpe:2017jej,Briceno:2018mlh,Briceno:2018aml,Blanton:2019igq,Briceno:2019muc,Romero-Lopez:2019qrt,Hansen:2020zhy,Meissner:2014dea,Blanton:2020jnm,Blanton:2020gha,Doring:2018xxx,Pang:2019dfe,Pang:2020pkl,Jackura:2019bmu,Meng:2017jgx,Mai:2018djl,Mai:2019fba,Konig:2017krd,Klos:2018sen,Konig:2020lzo,Blanton:2020gmf} (note also earlier
work on the related issues~\cite{Kreuzer:2010ti,Kreuzer:2009jp,Kreuzer:2008bi,Kreuzer:2012sr}). The reader is referred to~\cite{Hansen:2019nir} for a recent review on the three-particle problem in a finite volume. An alternative approach
to the problem focuses on the {\em perturbative} shift of the free
three-particle energy levels. Albeit more restricted from
the beginning (e.g., it is clear that the resonances cannot be handled in this
manner), the approach is very popular at present. The reason for this is that
the perturbative approach is very transparent and convenient for the analysis of
lattice data. In case of identical particles,
the expansion parameter is given by $a/L$, where $a$ denotes the
two-body scattering length and $L$ is the spatial size of the cubic box. Up
to and including $O(L^{-5})$, only two-body threshold parameters -- the scattering
length $a$ and the effective range $r$ -- enter the expression for the energy shift, and the latter appears at this order only in the excited states. At order $L^{-6}$, a single additional
coupling $\eta_3$, characterizing the three-body nonderivative interaction,
appears in the perturbative energy shift, and hence the measured three-body energies at different values of $L$ can be used to extract the value of this coupling from data. Moreover, the perturbative approach can be directly generalized for the systems of four and more particles (which is presently
not the case for the exact quantization condition), and the measurements
in the many-particle sector can be used to restrict the fit of $\eta_3$, since the same coupling appears in all $N$-particle energies up to and including $O(L^{-6})$. An example of such a fit is given in the recent article~\cite{Romero-Lopez:2020rdq}.

Application of the perturbation theory to the calculation of the energy shifts
in a finite volume has a decades-long history~\cite{Lee:1957zzb,Huang:1957im,Wu:1959zz,Tan:2007bg}. This development culminated in
Refs.~\cite{Beane:2007qr,Detmold:2008gh} where, using the nonrelativistic effective field theory, the $N$-particle ground-state
shift is derived up to and including order $L^{-6}$ and $L^{-7}$, respectively.
The relativistic corrections at $O(L^{-6})$ are calculated in the same approach
in Ref.~\cite{Beane:2020ycc} where, in addition, the inclusion of the electromagnetic interactions is considered. Further, carrying out the threshold expansion
in the nonrelativistic limit, the expression for the energy shift of the ground and first excited states is derived in Ref.~\cite{Pang:2019dfe} at $O(L^{-6})$.
On the other hand, the threshold
expansion of the relativistic quantization condition leads to the relativistic expression of the ground-state energy at the same order~\cite{Hansen:2016fzj}. Finally, in Ref.~\cite{Romero-Lopez:2020rdq}, where the nonrelativistic
framework has been used, 
the relativistic corrections in the ground and first excited states have
been calculated up to and including $O(L^{-6})$.

All above studies have one thing in common -- they are carried out for the system of $N$ identical particles. Taking into account that the simplest three-particle system, which can be simulated in lattice QCD, is that of three pions, one may conclude that the expressions, obtained in the above papers, apply to the
case with maximal isospin only. It is,
however, interesting to simulate systems with a different total isospin, in order to extract
all low-energy
three-pion couplings. In addition, studying the three-pion decays on the lattice,
one does not encounter, in general, a three-pion system with maximal isospin in the final
state.
A perturbative expression for the energy shift for $I\neq 3$
at the required accuracy is, however, unavailable in the literature (to the best of
our knowledge, a single study~\cite{Hansen:2020zhy}, which treats different
isospin channels, deals with an exact quantization condition and the threshold expansion is not performed there). Our present study aims at closing this gap. As already mentioned,
the results of this study will be certainly useful, for example,
in the analysis of lattice data on the three-pion decays, which are expected in the
near future.

The layout of the paper is as follows. In Sec.~\ref{sec:lagr} we write down
the nonrelativistic two- and three-body Lagrangians and perform the matching.
Further, in Sec.~\ref{sec:states} we construct the three-pion states with
different total isospin $I$ and calculate the potentials -- the matrix elements
of the interaction Hamiltonian between the three-pion states. In Sec.~\ref{sec:results}, we derive the general expression for the perturbative energy shift up to and including $O(L^{-6})$ in the state with
the total isospin $I=1$. Finally, Sec.~\ref{sec:concl} contains our conclusions.

\section{Non-relativistic effective Lagrangians and matching}
\label{sec:lagr}

Our calculations closely follow the path of
Refs.~\cite{Romero-Lopez:2020rdq,Beane:2007qr}, where the identical particles have been considered.
Throughout this paper, we shall use the nonrelativistic effective field theory framework,
including relativistic effects perturbatively. We further assume isospin symmetry and
denote the common mass of pions by $M$. The Lagrangian, which will be used in the
calculations up to and including order $L^{-6}$, is given by
\eq\label{eq:lagr}
{\cal L}=\sum_{i=\pm,0}\pi_i^\dagger\biggl(i\partial_t-M+\frac{\nabla^2}{2M}+\frac{\nabla^4}{8M^3}\biggr)\pi^{}_i+{\cal L}_2+{\cal L}_3\, .
\en
Here, $\pi_\pm$ and $\pi^{}_0$ denote nonrelativistic pion fields, and ${\cal L}_2$ and ${\cal L}_3$ are the Lagrangians in the two- and three-particle sectors, respectively. Note that only leading relativistic
corrections in the kinetic term contribute up to the order $L^{-6}$.

\medskip
 
Further, at the order we are working, it suffices to retain in ${\cal L}_2$ the terms up to
the two spatial
derivatives. Then, this Lagrangian takes the form
\eq
  \label{eqn:two-particle_lagrangian}
  {\cal L}_2 &= & \frac{1}{2}\, c_1 \pi_0^\dagger \pi_0^\dagger \pi^{}_0 \pi^{}_0
  - \frac{1}{8}\, d_1\biggl( \pi_0^\dagger \leftrightnablasquare \pi_0^\dagger \pi^{}_0 \pi^{}_0+\mbox{h.c.}\biggr)
\nonumber\\[2mm]
  &+& 2c_2 \biggl( \pi_+^\dagger \pi_0^\dagger \pi_
+ \pi^{}_0 + \pi_-^\dagger \pi_0^\dagger \pi^{}_- \pi^{}_0 \biggr)
- \frac{1}{2}d_2 \biggl( \pi_+^\dagger \leftrightnablasquare \pi_0^\dagger \pi^{}_+ \pi^{}_0
+ \pi_-^\dagger \leftrightnablasquare \pi_0^\dagger \pi^{}_- \pi^{}_0  + \mbox{h.c.} \biggr)
\nonumber\\[2mm]
&+& c_3 \biggl( \pi_+^\dagger \pi_-^\dagger \pi^{}_0 \pi^{}_0 + \text{h.c.} \biggr)
- \frac{1}{4}\,d_3 \biggl( \pi_+^\dagger \leftrightnablasquare \pi_-^\dagger \pi^{}_0 \pi^{}_0 + \pi_0^\dagger \leftrightnablasquare \pi_0^\dagger \pi^{}_+ \pi^{}_- + \mbox{h.c.} \biggr)
\nonumber\\[2mm]
& +& 2c_4  \pi_+^\dagger \pi_-^\dagger \pi^{}_+ \pi^{}_-
- \frac{1}{2}\,d_4\biggl( \pi_+^\dagger \leftrightnablasquare \pi_-^\dagger \pi^{}_+ \pi^{}_-
+\mbox{h.c.}\biggr)
\nonumber\\[2mm]
& +& \frac{1}{2}\,c_5 \biggl( \pi_+^\dagger \pi_+^\dagger \pi^{}_+ \pi^{}_+
+ \pi_-^\dagger \pi_-^\dagger \pi^{}_- \pi^{}_- \biggr)
- \frac{1}{8}\,d_5 \biggl( \pi_+^\dagger \leftrightnablasquare \pi_+^\dagger \pi^{}_+ \pi^{}_+
+ \pi_-^\dagger \leftrightnablasquare \pi_-^\dagger \pi^{}_- \pi^{}_-  +\mbox{h.c.}\biggr)\, .
\en
Here, $\leftrightnabla = (\rightnabla - \leftnabla)/2$ is the Galilean-invariant derivative.
Note also that there are more terms with two derivatives, which describe the
center-of-mass motion of two-particle pairs. Since these terms do not contribute to the
energy shift of the ground state we are interested in~\cite{Romero-Lopez:2020rdq},
we do not display them here explicitly. In addition, it can be checked that the P-wave
terms do not contribute to the ground-state energy at this order
(albeit they contribute to the excited states). These terms are omitted as well.

In a finite box, the Fourier transform of the field takes the form
\eq
\pi_i({\bf x},t)=\frac{1}{L^3}\,\sum_{\bf p}e^{-ip^0t+i{\bf p}{\bf x}}a_i({\bf p})\, ,
\en
where the creation and annihilation operators obey the commutation relations:
\eq
[a_i({\bf p}),a_j^\dagger({\bf q})]=L^3\delta_{ij}\delta_{{\bf p}{\bf q}}\, .
\en
The normalized one-particle states are given by
\eq
|\pi_i({\bf p})\rangle=\frac{1}{L^{3/2}}\,a^\dagger_i({\bf p})|0\rangle\, .
\en
If dimensional regularization is used, the couplings $c_i$ and $d_i$
can be related to the parameters of the effective-range expansion in the $\pi\pi$
scattering phase shift through the matching:
	\eq
        p \cot \delta_I(p) = -\frac{1}{a_I} + \frac{1}{2} r_I p^2 + \cdots,
        \quad\quad I=0,2\, .
	\en
        The matching
        condition takes the form
	\begin{IEEEeqnarray}{rClrCl}
		c_1 &=& -\frac{4\pi}{M} \frac{1}{3}(a_0 + 2a_2),\quad\quad\quad\quad	& d_1 &=& -\frac{4\pi}{M} \frac{1}{3}(a_0^2 \hat r^{}_0 + 2a_2^2 \hat r^{}_2)\, , \nonumber \\[2mm]
		c_2 &=& -\frac{4\pi}{M} \frac{1}{2}a_2,\quad & d_2 &=& -\frac{4\pi}{M} \frac{1}{2}a_2^2 \hat r^{}_2\, ,\nonumber  \\[2mm]
		c_3 &=& -\frac{4\pi}{M} \frac{1}{3}(a_2 - a_0),\quad & d_3 &=& -\frac{4\pi}{M} \frac{1}{3}(a_2^2 \hat r^{}_2 - a_0^2 \hat r^{}_0)\, , \nonumber  \\[2mm] 
		c_4 &=& -\frac{4\pi}{M} \frac{1}{6}(2a_0 + a_2),\quad & d_4 &=& -\frac{4\pi}{M} \frac{1}{6}(2a_0^2 \hat r^{}_0 + a_2^2 \hat r^{}_2)\, ,\nonumber  \\[2mm]
		c_5 &=& -\frac{4\pi}{M} a_2,\quad & d_5 &=& -\frac{4\pi}{M} a_2^2 \hat r^{}_2 \, .
              \end{IEEEeqnarray}
              In the above equations, the quantities $\hat r^{}_I$ include relativistic corrections, which stem from the matching to the relativistic amplitude~\cite{Romero-Lopez:2020rdq}:
              \eq
              \hat r^{}_I=r^{}_I-\frac{1}{a^{}_IM^2}\, .
              \en
Further, in the calculations up to and including $O(L^{-6})$, only the nonderivative
Lagrangians in the three-particle sector contribute. It is possible to construct two linearly
independent terms with this property and, hence, the Lagrangian can be written as
\eq
{\cal L}_3 =
\eta_1 \biggl( \pi_+^\dagger\pi^{}_+ + \pi_0^\dagger\pi^{}_0 + \pi_-^\dagger\pi^{}_- \biggr)^3 
+ \eta_2 \biggl( \pi_+^\dagger\pi^{}_+ + \pi_0^\dagger\pi^{}_0 + \pi_-^\dagger\pi^{}_- \biggr)
\biggl( 2 \pi_+^\dagger \pi_-^\dagger - \pi_0^\dagger\pi_0^\dagger \biggr)
\biggl( 2 \pi^{}_+ \pi^{}_- - \pi^{}_0\pi^{}_0 \biggr)\,.
	\en
The low-energy couplings $\eta_1$ and $\eta_2$ are ultraviolet-divergent -- these divergences
cancel the pertinent divergences, which arise in the perturbation theory at higher orders.
We shall see the examples of such cancellations below.
Note also that $\eta_1$ is a counterpart of the coupling $\eta_3$, introduced
in Ref.~\cite{Beane:2007qr} for the case of three identical particles. Namely, one
has  $\eta_3=-6\eta_1$. On the contrary, the coupling $\eta_2$ does not emerge in the
sector with maximal isospin. In order to fix this coupling, one has to measure the
energy shift in the sector with $I=1$.

\medskip

It is well known that the couplings
$\eta_1$ and $\eta_2$, via matching, can be traded for the ``divergence free'' threshold
amplitudes in the three-particle sector.
Such an amplitude can be defined in different ways. In the present context, the
definition, introduced in Ref.~\cite{Romero-Lopez:2020rdq}, is the most convenient,
and we stick to it. In order to match two constants, we consider two relativistic on-shell
amplitudes $3\pi^+\to 3\pi^+$ and $3\pi^0\to 3\pi^0$, denoted by ${\cal M}_+$ and
${\cal M}_0$, respectively. Next, we have to choose the
directions from which the threshold is approached. This can be done, e.g., in the
following way:
\eq
{\bf p}_1=\lambda{\bf e}_y\, ,\quad\quad
{\bf p}_2=\lambda\biggl(\frac{\sqrt{3}}{2}\,{\bf e}_x-\frac{1}{2}\,{\bf e}_y\biggr)\, ,\quad\quad
{\bf p}_3=-\lambda\biggl(\frac{\sqrt{3}}{2}\,{\bf e}_x+\frac{1}{2}\,{\bf e}_y\biggr)\, ,
\en
and ${\bf p}_i'=-{\bf p}_i$ for $i=1,2,3$. Here, ${\bf p}_i,{\bf p}'_i$ are the initial
and final momenta of particles and ${\bf e}_{x,y,z}$ denote unit vectors in the direction
of pertinent axes.

The limit $\lambda\to 0$ in the above amplitudes is singular. In order to arrive at the
regular quantities, one has to subtract the one-particle irreducible piece (pole part) first.
An exact definition of this quantity can be found in Ref.~\cite{Romero-Lopez:2020rdq}.
Then, it can be proven that
\eq
\mbox{Re}\biggl({\cal M}_a-{\cal M}_a^{(pole)}\biggr)
=\frac{1}{\lambda}\,{\cal M}_a^{(-1)}+\ln\frac{\lambda}{M}\,{\cal M}_a^{(l)}
+{\cal M}_a^{(0)}+\cdots\, ,\quad\quad a=+,0\, .
\en
Here, ellipses stand for the terms that vanish, as $\lambda\to 0$.

\medskip

The couplings $\eta_1$ and $\eta_2$ can be related to the regular threshold amplitudes
${\cal M}_a^{(0)}$. This relation can be obtained exactly in the same way as
in Ref.~\cite{Romero-Lopez:2020rdq}, and we quote here only the final result:
\eq\label{eq:matching-eta}
36\eta_1^r&=&-\frac{288\pi^2}{M}\,a_2^3\hat r_2-\frac{576\pi^2}{M^3}\,a_2^2
+\frac{1}{(2M)^3}\,{\cal M}_+^{(0)}-\frac{192\pi a_2^4}{M}\,\bigl(3\sqrt{3}\bar\delta^{(d)}-4\pi\bar\delta^{(e)}\bigr)\, ,
\nonumber\\[2mm]
36(\eta_1^r+\eta_2^r)&=&-\frac{32\pi^2 }{M}\,(a_0+2a_2)(a_0^2\hat r_0+2a_2^2\hat r_2)
-\frac{64\pi^2}{M^3}(a_0+2a_2)^2
+\frac{1}{(2M)^3}\,{\cal M}_0^{(0)}
\nonumber\\[2mm]
&-&\frac{64\sqrt{3}\pi}{3M}\,(a_0^4+4a_0^3a_2+9a_0^2a_2^2+2a_0a_2^3+11a_2^4)
\bar\delta^{(d)}
\nonumber\\[2mm]
&+&\frac{128\pi^2}{27M}\,(2a_0^4+28a_0^3a_2+57a_0^2a_2^2+10a_0a_2^3+65a_2^4)\bar\delta^{(e)}\, .
\en
        Here, $\eta_{1,2}^r$ denote the finite parts of the couplings $\eta_{1,2}$
        (dimensional regularization with the minimal subtraction prescription has been used,
        and $d$ denotes the number of space dimensions):
        \eq
        \eta_1&=&\mu^{2(d-3)}\biggl\{-\frac{16\pi a_2^4}{3M}\,\frac{1}{d-3}\,(3\sqrt{3}-4\pi)+\eta_1^r\biggr\}\, ,
        \nonumber\\[2mm]
        \eta_2&=&\mu^{2(d-3)}\biggl\{
        \frac{16\pi}{243M}\,\frac{1}{d-3}\,(a_0-a_2)\Big[ 2\pi	\left(2a_0^3 + 30 a_0^2 a_2 + 87 a_0 a_2^2 +97 a_2^3\right) - 9\sqrt{3}\left(a_0 + 2a_2\right)\left(a_0^2 + 3a_0a_2 + 8a_2^2\right) \Big]
\nonumber\\[2mm]
        &+&\eta_2^r\biggr\}\, .
        \en
        Furthermore, the renormalization scale was chosen at $\mu=M$ in the above
        equations.

\medskip

The quantities $\bar\delta^{(d)}$ and $\bar\delta^{(e)}$, which emerge in
Eq.~(\ref{eq:matching-eta}), are equal to
\eq
\bar\delta^{(d,e)}=\delta^{(d,e)}-\Gamma'(1)-\ln 4\pi\, ,\quad\quad
\delta^{(d)}=-1.090\ldots\, ,\quad\quad
\delta^{(e)}=3.926\ldots\, .
\en
where $\delta^{(d)}$ and $\delta^{(e)}$ are the quantities defined in Appendix A of
Ref.~\cite{Romero-Lopez:2020rdq}.

\medskip

To summarize, one can express two unknown couplings $\eta_1^r$ and $\eta_2^r$ through the
regular, real-valued threshold amplitudes ${\cal M}_+^{(0)}$ and ${\cal M}_0^{(0)}$, which
can be further extracted from the measured energy levels on the lattice. At this stage,
we do not see much advantage in
working in terms of the amplitudes rather than couplings,
since this amounts to the replacing of one set of
unknown constants by another one in the fit.
Note, however, that the amplitudes are observables,
whereas the couplings are not. For this reason, comparison with other approaches is
made easier, if the final result is rewritten in terms of the amplitudes.

\section{The Fock state and the potentials}
\label{sec:states}
In a finite cubic box with size $L$, the Fock state of the theory consists of the vectors, containing freely moving particles $\pi_\pm$ and $\pi^{}_0$ with the discretized momenta ${\bf p}=2\pi{\bf n}/L$, where ${\bf n}\in\mathbb{Z}^3$. For instance, the closure relation in the three-particle sector
with a total charge +1 is written in the following form:
\eq\label{eq:closure}
\mathbbm{1}&=&\frac{1}{2!}\,\sum_{{\bf p}_1,{\bf p}_2,{\bf p}_3}
|\pi_+({\bf p}_1)\pi_+({\bf p}_2)\pi_-({\bf p}_3)\rangle
\langle \pi_+({\bf p}_1)\pi_+({\bf p}_2)\pi_-({\bf p}_3)|
\nonumber\\[2mm]
&+&\frac{1}{2!}\,\sum_{{\bf p}_1,{\bf p}_2,{\bf p}_3}
|\pi_0({\bf p}_1)\pi_0({\bf p}_2)\pi_+({\bf p}_3)\rangle
\langle \pi_0({\bf p}_1)\pi_0({\bf p}_2)\pi_+({\bf p}_3)|\, .
\en
Note that here the ``odd'' particle is always labeled by the index $3$. Instead, one could
sum up over three permutations of indices $(123)$, $(312)$, $(231)$, in each of the above sums,
dividing everything by $3!$ instead of $2!$.

\medskip
 
Next, we wish to rewrite this sum as a sum over the three-pion states with a definite
total isospin $I$. It is a straightforward exercise to construct such states, using the table
of the Clebsh-Gordan coefficients. Below, we give maximal vectors $I_z=I$ in each
of the representation with $I=0,1,2,3$:
\begin{IEEEeqnarray}{lCl}\label{eqn:isospin_decompositon_first}
  |3,3\rangle &=& |\pi_+ \pi_+ \pi_+\rangle\, , \nonumber\\ 
			\phantom{} \nonumber\\
			|2,2\rangle_A &=& \frac{1}{\sqrt{2}}\,
             \biggl( |\pi_+\pi_0\pi_+\rangle - |\pi_0\pi_+\pi_+\rangle \biggr)\, , \nonumber\\[2mm]
			|2,2\rangle_S &=& \frac{1}{\sqrt{6}}\,
                        \biggl( 2|\pi_+\pi_+\pi_0\rangle -  |\pi_0\pi_+\pi_+\rangle -
                        |\pi_+\pi_0\pi_+\rangle  \biggr)\, ,  \nonumber\\
			\phantom{} \nonumber\\ 
			|1,1\rangle_A &=& \frac{1}{2}\, \biggl( |\pi_+\pi_0\pi_0\rangle - |\pi_0\pi_+\pi_0\rangle - |\pi_+\pi_-\pi_+\rangle + |\pi_-\pi_+\pi_+\rangle \biggr)\, , \nonumber\\[2mm]  
			|1,1\rangle_{S_1} &=& \frac{1}{\sqrt{3}}\, \biggl( |\pi_+\pi_-\pi_+\rangle - |\pi_0\pi_0\pi_+\rangle + |\pi_-\pi_+\pi_+\rangle \biggr)\, ,\nonumber \\[2mm]
                        |1,1\rangle_{S_2} &=&  \frac{1}{2\sqrt{15}}\,
                        \biggl(  6|\pi_+\pi_+\pi_-\rangle
                        +  |\pi_+\pi_-\pi_+\rangle+ |\pi_-\pi_+\pi_+\rangle
                        - 3|\pi_0\pi_+\pi_0\rangle -3|\pi_+\pi_0\pi_0\rangle
	+ 2|\pi_0\pi_0\pi_+\rangle  \biggr)\, ,
		\nonumber\\
			\phantom{} \nonumber\\
	|0,0\rangle &=&\frac{1}{\sqrt{6}}\, \biggl( |\pi_+\pi_0\pi_-\rangle - |\pi_+\pi_-\pi_0\rangle - |\pi_0\pi_+\pi_-\rangle + |\pi_0\pi_-\pi_+\rangle
        + |\pi_-\pi_+\pi_0\rangle - |\pi_-\pi_0\pi_+\rangle \biggr)\, .
      	\end{IEEEeqnarray}
        Here, it is implicitly assumed that the particles in the ket-vectors carry momenta
        ${\bf p}_1,{\bf p}_2,{\bf p}_3$, respectively, and the subscripts $S/A$ stand for symmetric/antisymmetric wave functions with respect to $1\leftrightarrow 2$. Further, 
        the multiplicity of the irreducible representations, corresponding to $I=0,1,2,3$
        is $1,3,2,1$, respectively, i.e., the unperturbed states in the sectors $I=1,2$ are,
        generally, degenerate, and hence the perturbation theory for the degenerate states should be used in these sectors. Note also that this basis is the counterpart of the basis,
        used in Ref.~\cite{Hansen:2020zhy}, Eqs. (2.56) and (2.57) (there, $I_z=0$
        for all $I$ has been chosen).

\medskip
 
Let us now focus on the unperturbed ground state with ${\bf p}_1={\bf p}_2={\bf p}_3=0$.
From Eq.~(\ref{eqn:isospin_decompositon_first}) it is clear that only the vectors
with $I=3$ and $I=1$ survive, whereas for $I=0,2$, the unperturbed ground-state
vector vanishes identically -- these states will not be seen on the lattice.
Further, note that the solution for $I=3$ is
already known in the literature, whereas the expression for the shift in the $I=1$ state,
which will be given below, is new. Providing this expression,
our formalism that describes the ground state of three pions will become complete.

\medskip

At the next step, we wish to rewrite the closure relation in terms of the states with
a definite total isospin. We shall work with the full basis of states with $I_z=+1$, where
the states with $I=1$ are also contained. In the Fock space, there are six basis vectors
with $I_z=+1$:
        \begin{IEEEeqnarray}{rCrrCrrCr}
        |v_1\rangle&=&|\pi_+\pi_+\pi_-\rangle\, ,\quad\quad
        |v_2\rangle&=&|\pi_+\pi_-\pi_+\rangle\, ,\quad\quad
        |v_3\rangle&=&|\pi_-\pi_+\pi_+\rangle\, ,\quad\quad
        \nonumber\\[2mm]
        |v_4\rangle&=&|\pi_0\pi_0\pi_+\rangle\, ,\quad\quad
        |v_5\rangle&=&|\pi_0\pi_+\pi_0\rangle\, ,\quad\quad
        |v_6\rangle&=&|\pi_+\pi_0\pi_0\rangle\, .\quad\quad
\end{IEEEeqnarray}
We wish to express these vectors, as linear combinations of the vectors $|I,I_z\rangle$
        with $I=1,2,3$ and $I_z=1$. The explicit expression of these vectors is given by:
        \eq\label{eq:isospinbasis}
        |f_1\rangle&=& \frac{1}{\sqrt{15}}\, \biggl(  2|\pi_+\pi_+\pi_-\rangle
        +2|\pi_+\pi_-\pi_+\rangle +2|\pi_-\pi_+\pi_+\rangle
        - |\pi_0\pi_0\pi_+\rangle   - |\pi_0\pi_+\pi_0\rangle - |\pi_+\pi_0\pi_0\rangle\biggr)\, ,
        \nonumber \\[2mm]
                        |f_2\rangle&=& \frac{1}{2\sqrt{3}}\,
                        \biggl(  2|\pi_+\pi_+\pi_-\rangle
                        -  |\pi_+\pi_-\pi_+\rangle- |\pi_-\pi_+\pi_+\rangle
                        + 2|\pi_0\pi_0\pi_+\rangle   - |\pi_0\pi_+\pi_0\rangle -|\pi_+\pi_0\pi_0\rangle
                        \biggr)\, ,
                        \nonumber\\[2mm]
|f_3\rangle&=& \frac{1}{2}\, \biggl( |\pi_+\pi_0\pi_0\rangle - |\pi_0\pi_+\pi_0\rangle - |\pi_+\pi_-\pi_+\rangle + |\pi_-\pi_+\pi_+\rangle \biggr)\, , \nonumber\\[2mm]  
        |f_4\rangle&=&\frac{1}{2\sqrt{3}}\,\biggl(
 2|\pi_+\pi_+\pi_-\rangle-|\pi_+\pi_-\pi_+\rangle-|\pi_-\pi_+\pi_+\rangle
        -2|\pi_0\pi_0\pi_+\rangle+|\pi_0\pi_+\pi_0\rangle+|\pi_+\pi_0\pi_0\rangle\biggr)\, ,
        \nonumber\\[2mm]
|f_5\rangle&=&\frac{1}{2}\,\biggl(
          |\pi_+\pi_0\pi_0\rangle-|\pi_0\pi_+\pi_0\rangle+|\pi_+\pi_-\pi_+\rangle-|\pi_-\pi_+\pi_+\rangle\biggr)\, ,
        \nonumber\\[2mm]
      |f_6\rangle&=&\frac{1}{\sqrt{15}}\,\biggl(
        |\pi_+\pi_+\pi_-\rangle+|\pi_+\pi_-\pi_+\rangle+|\pi_-\pi_+\pi_+\rangle
        +2|\pi_0\pi_0\pi_+\rangle+2|\pi_0\pi_+\pi_0\rangle+2|\pi_+\pi_0\pi_0\rangle\biggr)\, .
        \en
        Note that the vectors $|f_1\rangle$ and $|f_2\rangle$ are linear combinations of the
        vectors $|1,1\rangle_{S_1}$ and $|1,1\rangle_{S_2}$ (the same linear combinations, as
        in Ref.~\cite{Hansen:2020zhy}, Eqs.~(C.14) and (C.17), again for $I_z=0$).
        The reason for such choice will become clear below.

\medskip
        
          It is now straightforwardly seen that the basis vectors of one basis can be expressed as linear
          combinations of the basis vectors of other basis: $|v_a\rangle=C_{ab}|f_b\rangle$.
          The orthogonal $6\times 6$ matrix $C$ is given by:
          \eq
          C=\begin{pmatrix}
\frac{2}{\sqrt{15}} & \frac{1}{\sqrt{3}} & 0 & \frac{1}{\sqrt{3}} & 0 &\frac{1}{\sqrt{15}} \\[2mm]
-\frac{2}{\sqrt{15}} & \frac{1}{2\sqrt{3}} &\frac{1}{2} &  \frac{1}{2\sqrt{3}} &-\frac{1}{2} &  -\frac{1}{\sqrt{15}} \\[2mm]
-\frac{2}{\sqrt{15}} & \frac{1}{2\sqrt{3}} &-\frac{1}{2} &  \frac{1}{2\sqrt{3}} &\frac{1}{2} &  -\frac{1}{\sqrt{15}} \\[2mm]
-\frac{1}{\sqrt{15}} & \frac{1}{\sqrt{3}} & 0 & -\frac{1}{\sqrt{3}} &0 & \frac{2}{\sqrt{15}} \\[2mm]
\frac{1}{\sqrt{15}} & \frac{1}{2\sqrt{3}} &\frac{1}{2} & -\frac{1}{2\sqrt{3}} &\frac{1}{2} &-\frac{2}{\sqrt{15}}  \\[2mm]
\frac{1}{\sqrt{15}} & \frac{1}{2\sqrt{3}} &-\frac{1}{2} & -\frac{1}{2\sqrt{3}} &-\frac{1}{2} &-\frac{2}{\sqrt{15}}
          \end{pmatrix}\, .
          \en
          Using $CC^T=C^TC=1$, the closure relation (\ref{eq:closure}) can be finally rewritten in the isospin basis:
          \eq
          \mathbbm{1}=\frac{1}{3!}\sum_{a=1}^6\sum_{{\bf p}_1{\bf p}_2{\bf p}_3}
          |f_a\rangle\langle f_a|\, .
          \en
          The Lippmann-Schwinger equation, which can be obtained in the nonrelativistic effective theory, can be partially diagonalized in this basis. In order to derive this equation, we write down the canonical Hamiltonian, obtained from Eq.~(\ref{eq:lagr}), in the following form:
          \eq
          {\bf H}={\bf H}_0+{\bf H}_I={\bf H}_0+{\bf H}_c+{\bf H}_d+{\bf H}_r+{\bf H}_\eta\, .
          \en
          Here ${\bf H}_0$ is the free nonrelativistic Hamiltonian. The states, which were constructed above, are the eigenstates of ${\bf H}_0$:
          \eq
          {\bf H}_0|\pi_i({\bf p}_1)\pi_j({\bf p}_2)\pi_k({\bf p}_3)\rangle
          &=&\frac{1}{2M}\,({\bf p}_1^2+{\bf p}_2^2+{\bf p}_3^2)|\pi_i({\bf p}_1)\pi_j({\bf p}_2)\pi_k({\bf p}_3)\rangle
\nonumber\\[2mm]
          &\doteq& E_p|\pi_i({\bf p}_1)\pi_j({\bf p}_2)\pi_k({\bf p}_3)\rangle\, ,\quad\quad i,j,k=\pm,0\, .
          \en
          Further, various terms in the interaction Lagrangian are given by
\eq
          {\bf H}_c&=&\int d^3{\bf x}\biggl\{
          -\frac{1}{2}\, c_1 \pi_0^\dagger \pi_0^\dagger \pi^{}_0 \pi^{}_0
          -2c_2 \biggl( \pi_+^\dagger \pi_0^\dagger \pi^{}_+ \pi^{}_0
          + \pi_-^\dagger \pi_0^\dagger \pi^{}_- \pi^{}_0 \biggr)
- c_3 \biggl( \pi_+^\dagger \pi_-^\dagger \pi^{}_0 \pi^{}_0 + \text{h.c.} \biggr)
- 2c_4  \pi_+^\dagger \pi_-^\dagger \pi^{}_+ \pi^{}_-
\nonumber\\[2mm]
&-&\frac{1}{2}\,c_5 \biggl( \pi_+^\dagger \pi_+^\dagger \pi^{}_+ \pi^{}_+
+ \pi_-^\dagger \pi_-^\dagger \pi^{}_- \pi^{}_- \biggr)\biggr\}\, ,
\\[2mm]
{\bf H}_d&=&\int d^3{\bf x}\biggl\{
  \frac{1}{8}\, d_1\biggl( \pi_0^\dagger \leftrightnablasquare \pi_0^\dagger \pi^{}_0 \pi^{}_0+\mbox{h.c.}\biggr)
 + \frac{1}{2}d_2 \biggl( \pi_+^\dagger \leftrightnablasquare \pi_0^\dagger \pi^{}_+ \pi^{}_0
+ \pi_-^\dagger \leftrightnablasquare \pi_0^\dagger \pi^{}_- \pi^{}_0  + \mbox{h.c.} \biggr)
\nonumber\\[2mm]
&+& \frac{1}{4}\,d_3 \biggl( \pi_+^\dagger \leftrightnablasquare \pi_-^\dagger \pi_0 \pi^{}_0 + \pi_0^\dagger \leftrightnablasquare \pi_0^\dagger \pi^{}_+ \pi^{}_- + \mbox{h.c.} \biggr)
+ \frac{1}{2}\,d_4\biggl( \pi_+^\dagger \leftrightnablasquare \pi_-^\dagger \pi^{}_+ \pi^{}_-
+\mbox{h.c.}\biggr)
\nonumber\\[2mm]
&+& \frac{1}{8}\,d_5 \biggl( \pi_+^\dagger \leftrightnablasquare \pi_+^\dagger \pi^{}_+ \pi^{}_+
+ \pi_-^\dagger \leftrightnablasquare \pi_-^\dagger \pi^{}_- \pi^{}_-  +\mbox{h.c.}\biggr)\biggr\}\, ,
\\[2mm]
{\bf H}_r&=&-\frac{1}{8M^3}\sum_{i=\pm,0}
\int d^3{\bf x}\biggl\{\pi_i^\dagger\nabla^4\pi^{}_i\biggr\}\, ,
\\[2mm]
{\bf H}_\eta &=&\int d^3{\bf x}\biggl\{
-\eta_1 \biggl( \pi_+^\dagger\pi^{}_+ + \pi_0^\dagger\pi^{}_0 + \pi_-^\dagger\pi^{}_- \biggr)^3 
- \eta_2 \biggl( \pi_+^\dagger\pi^{}_+ + \pi_0^\dagger\pi^{}_0 + \pi_-^\dagger\pi^{}_- \biggr)
\biggl( 2 \pi_+^\dagger \pi_-^\dagger - \pi_0^\dagger\pi_0^\dagger \biggr)
\biggl( 2 \pi^{}_+ \pi^{}_- - \pi^{}_0\pi^{}_0 \biggr)\biggr\}\,.\nonumber\\
\en
The integration in the above expressions is carried out in a cubic box with a size $L$.

\medskip

In the operator form, the Lippmann-Schwinger equation can be written as
\eq
{\bf T}(E)={\bf H}_I+{\bf H}_I{\bf G}_0(E){\bf T}(E)\, ,\quad\quad
{\bf G}_0(E)=\frac{1}{E-{\bf H}_0}\, .
\en
Let us now sandwich this equation by the three-particle states with the definite total isospin
$I$, use closure relations and take into account the fact that ${\bf T}$, ${\bf H}_I$ and ${\bf G}_0$ are diagonal in $I$. In particular, we are primarily interested in the sector $I=1$,
where the equation simplifies to
\eq\label{eq:LS}
\langle p|{\bf T}_{\alpha\beta}(E)|q\rangle
=\langle p|{\bf V}_{\alpha\beta}|q\rangle
+\sum_k\sum_{\gamma=1}^3\langle p|{\bf V}_{\alpha\gamma}|k\rangle\frac{1}{E-E_k}\,
\langle k|{\bf T}_{\gamma\beta}(E)|q\rangle\, ,
\en
where, the symbol $p$ denotes a set of momenta $({\bf p}_1,{\bf p}_2,{\bf p}_3)$, and so on. Further, the matrix elements are defined as
\eq
\langle p|{\bf V}_{\alpha\beta}|q\rangle=\frac{1}{3!}\,
\langle f_\alpha,p|{\bf H}_I|f_\beta,q\rangle\, ,\quad\quad
\langle p|{\bf T}_{\alpha\beta}(E)|q\rangle=\frac{1}{3!}\,
\langle f_\alpha,p|{\bf T}(E)|f_\beta,q\rangle\, .
\en
Here, in difference with the indices $a$ and $b$, the indices $\alpha,\beta,\gamma$
run from 1 to 3, picking out the vectors $f_1,f_2,f_3$ from the sector $I=1$ only. Thus, the partial
diagonalization
of the Lippmann-Schwinger equation is achieved in the basis of the isospin
vectors~(\ref{eq:isospinbasis}).

\medskip

The matrix elements of the interaction Hamiltonian in the above equation can be calculated
straightforwardly. Below, we merely list different contributions to this matrix element
in the sector with $I=1$:
\eq \label{eqn:I=1_two-body-potential_first}
\langle f_1,p|{\bf H}_c|f_1,q\rangle &=& \frac{4\pi}{M L^3} \frac{1}{27} (5a_0 +4a_2)\biggl[ V_1 + V_2 + V_3 + V^+_1 + V^+_2 +  V^+_3\biggr]\, , \nonumber\\[2mm]
\langle f_1,p|{\bf H}_c|f_2,q\rangle &=& \frac{4\pi}{M L^3} \frac{\sqrt{5}}{54} (a_0 - a_2)\biggl[ 2\left(V_1 + V_2 -2V_3\right) - 3\left(V^-_1+V^-_2\right) - \left(V^+_1+V^+_2-2V^+_3\right) \biggr]\, , \nonumber\\[2mm]
	\langle f_2,p|{\bf H}_c|f_2,q\rangle &=& \frac{4\pi}{M L^3} \frac{1}{108} (4a_0 + 5a_2)\biggl[ \left(V_1 + V_2 + 4V_3\right) -\left(2V^+_1+2V^+_2-V^+_3\right) \biggr]\, , \nonumber\\[2mm]
        \langle f_1,p|{\bf H}_c|f_3,q\rangle &=& -\frac{4\pi}{M L^3} \frac{\sqrt{5}}{18\sqrt{3}} (a_0 - a_2)\biggl[ 2\left(V_1-V_2\right) + \left(V^-_1-V^-_2-2V^-_3\right) - \left(V^+_1 - V^+_2\right)\biggr]\, , \nonumber\\[2mm]
        \langle f_2,p|{\bf H}_c|f_3,q\rangle &=& -\frac{4\pi}{M L^3} \frac{1}{36\sqrt{3}} (4a_0 + 5a_2)\biggl[ \left(V_1 - V_2\right) -\left(V^-_1 - V^-_2+V^-_3\right) + \left(V^+_1-V^+_2\right)\biggr] \, ,\nonumber\\[2mm]
\langle f_3,p|{\bf H}_c|f_3,q\rangle  &=& \frac{4\pi}{M L^3} \frac{1}{36} (4a_0+5a_2)
(V_1+V_2-V^+_3)\, ,
                        \en
                        where
\eq\label{eq:V}
			V_1 = \delta_{\mathbf{p}_1,\mathbf{k}_1}\delta_{\mathbf{p}_2+\mathbf{p}_3,\mathbf{k}_2+\mathbf{k}_3}, \quad\quad
			V_2 = \delta_{\mathbf{p}_2,\mathbf{k}_2}\delta_{\mathbf{p}_1+\mathbf{p}_3,\mathbf{k}_1+\mathbf{k}_3}, \quad\quad
			V_3 = \delta_{\mathbf{p}_3,\mathbf{k}_3}\delta_{\mathbf{p}_1+\mathbf{p}_2,\mathbf{k}_1+\mathbf{k}_2},
	\en
	and
	\eq\label{eq:hatV}
		V^\pm_1 &=& \delta_{\mathbf{p}_2,\mathbf{k}_3} \delta_{\mathbf{p}_1+\mathbf{p}_3,\mathbf{k}_1+\mathbf{k}_2} \pm \delta_{\mathbf{p}_3,\mathbf{k}_2} \delta_{\mathbf{p}_1+\mathbf{p}_2,\mathbf{k}_1+\mathbf{k}_3}\, , \nonumber\\[2mm]
		V^\pm_2&=& \delta_{\mathbf{p}_1,\mathbf{k}_3} \delta_{\mathbf{p}_2+\mathbf{p}_3,\mathbf{k}_1+\mathbf{k}_2} \pm \delta_{\mathbf{p}_3,\mathbf{k}_1} \delta_{\mathbf{p}_1+\mathbf{p}_2,\mathbf{k}_2+\mathbf{k}_3}\, ,\nonumber\\[2mm]
		V^\pm_3 &=& \delta_{\mathbf{p}_1,\mathbf{k}_2} \delta_{\mathbf{p}_2+\mathbf{p}_3,\mathbf{k}_1+\mathbf{k}_3} \pm \delta_{\mathbf{p}_2,\mathbf{k}_1} \delta_{\mathbf{p}_1+\mathbf{p}_3,\mathbf{k}_2+\mathbf{k}_3}\, .
              \en
              Further, the matrix elements of the operator ${\bf H}_d$ are also given  by
              Eq.~(\ref{eqn:I=1_two-body-potential_first}), with the replacement
              $a_I^{}\to a_I^{2}\hat r_I$ and $V^{}_i,V_i^\pm\to \hat V^{}_i,\hat V_i^\pm$. The latter are again given by
              Eqs.~(\ref{eq:V}) and (\ref{eq:hatV}), with the following replacements done:
	\eq
	 \delta_{\mathbf{p}_i,\mathbf{k}_l}\delta_{\mathbf{p}_j+\mathbf{p}_k,\mathbf{k}_m+\mathbf{k}_n} \rightarrow \frac{1}{16}\, \delta_{\mathbf{p}_i,\mathbf{k}_l}\delta_{\mathbf{p}_j+\mathbf{p}_k,\mathbf{k}_m+\mathbf{k}_n} \left[ (\mathbf{p}_j - \mathbf{p}_k )^2 + (\mathbf{k}_m - \mathbf{k}_n )^2 \right].
	\en
        Next, the relativistic correction in the kinetic term has
        the following matrix elements:
        	\eq
                \langle f_1,p|{\bf H}_r|f_1,q\rangle &=& -\frac{1}{96 M^3}
                \sum_{i=1}^3 (\mathbf{k}^4_i + \mathbf{p}^4_i) \biggl( V_{123} + V_{132} + V_{231}  + V_{213} + V_{312} + V_{321} \biggr)\, , \nonumber\\[2mm]
\langle f_1,p|{\bf H}_r|f_2,q\rangle &=& 0\, ,\nonumber\\[2mm]
\langle f_2,p|{\bf H}_r|f_2,q\rangle &=& -\frac{1}{192M^3} \sum_{i=1}^3( \mathbf{k}^4_i + \mathbf{p}^4_i) \biggl( 2 V_{123} - V_{132} + 2V_{213}  - V_{231} - V_{312} - V_{321} \biggr)\, , \nonumber\\[2mm]
\langle f_1,p|{\bf H}_r|f_3,q\rangle &=& 0\, ,\nonumber\\[2mm]
\langle f_2,p|{\bf H}_r|f_3,q\rangle &=& \frac{1}{64\sqrt{3}M^3} \sum_{i=1}^3( \mathbf{k}^4_i + \mathbf{p}^4_i) \biggl( V_{132}  - V_{231} + V_{312} - V_{321} \biggr)\, , \nonumber\\[2mm]
\langle f_3,p|{\bf H}_r|f_3,q\rangle &=& -\frac{1}{192 M^3} \sum_{i=1}^3( \mathbf{k}^4_i + \mathbf{p}^4_i) \biggl( 2 V_{123} - 2V_{213} + V_{132}  - V_{231} - V_{312} + V_{321} \biggr)\, ,
\en
where
	\eq
		 V_{ijk} = \delta_{\mathbf{p}_1,\mathbf{k}_i} \delta_{\mathbf{p}_2,\mathbf{k}_j} \delta_{\mathbf{p}_3,\mathbf{k}_k}\, .
	\en
        Finally, the single non-zero matrix element of the three-particle Hamiltonian is given by:
        \eq
 \langle f_1,p|{\bf H}_\eta|f_1,q\rangle &=& 
 -\frac{2}{L^6}  \left(3\eta_1+5\eta_2\right)\, \delta_{\mathbf{p}_1+\mathbf{p}_2+\mathbf{p}_3,\mathbf{k}_1+\mathbf{k}_2+\mathbf{k}_3}\, .
  \en

 \section{Perturbation theory}
 \label{sec:results}

 The energy level shift in the nonrelativistic effective field theory can be calculated by using the
 ordinary Rayleigh-Schr\"odinger perturbation theory. A small technical issue might arise here since,
 as we have seen, there are multiple states with $I=1$, so, the application of the perturbation
 theory for the degenerate states might be needed. This is, however, not the case for the
 ground state -- as seen from Eq.~(\ref{eq:isospinbasis}), only the state $f_1$ has a nonvanishing
 ground-state wave function. Note that this is different in the excited states -- in this case,
 the perturbation theory for the degenerate states should indeed be used.
 
\medskip

In order to derive the perturbative expansion for the ground-state energy,
at the first step we single out the contribution from the
unperturbed ground-state level ${\bf k}_1={\bf k}_2={\bf k}_3=0$.
It can be seen that Eq.~(\ref{eq:LS}) is equivalent to the
 following system of linear equations:
 \eq\label{eq:LS1}
 \langle p|{\bf T}_{\alpha\beta}(E)|q\rangle&=&\langle p|\boldsymbol{\Omega}_{\alpha\beta}(E)|q\rangle
 +\sum_{\gamma=1}^3\langle p|\boldsymbol{\Omega}_{\alpha\gamma}(E)|0\rangle
 \frac{1}{E-E_0} \langle 0|{\bf T}_{\gamma\beta}(E)|q\rangle\, ,
\\[2mm]
\label{eq:LS2}
\langle p|\boldsymbol{\Omega}_{\alpha\beta}(E)|q\rangle
&=&\langle p|{\bf V}_{\alpha\beta}|q\rangle
+\sum_{k\neq 0}\sum_{\gamma=1}^3\langle p|{\bf V}_{\alpha\gamma}|k\rangle\frac{1}{E-E_k}\,
\langle k|\boldsymbol{\Omega}_{\gamma\beta}(E)|q\rangle\, .
\en
The shifted energy is given by the position of the pole of the scattering matrix.
Setting now external momenta to zero, $p=q=0$,
it is seen using Eq.~(\ref{eq:LS1}) that the shifted ground-state pole position obeys
the secular equation,
 \eq\label{eq:qc}
E-E_0-\hat\Omega(E)=0\, ,
 \en
 where $\hat\Omega(E) =\Omega_{11}(E)$ and $\Omega(E)$ is a $3\times 3$ matrix, whose components coincide
 with $\langle 0|\boldsymbol{\Omega}_{\alpha\beta}(E)|0\rangle$ (we remind the reader that
 all components of this matrix, except $\alpha=\beta=1$, vanish at $p=q=0$).

 \medskip

 At the next step, we recall that the quantity
 $\hat\Omega(E)$ does not contain the singular denominator $1/(E-E_0)$, which is excluded
 in the sum with $k\neq 0$.
 Consequently, one may expand $\hat\Omega(E)$ in Taylor series in the vicinity of the unperturbed ground state,
 \eq
 \hat\Omega(E)=\hat\Omega(E_0)+(E-E_0)\hat\Omega'(E_0)+\frac{1}{2}\,(E-E_0)^2\hat\Omega''(E_0)
 +O((E-E_0)^3)\, .
 \en
The coefficients of this expansion can be further expanded in the inverse powers of $L$:
	\eq
			\hat\Omega(E_0) &=& \frac{\hat\Omega_{3}(E_0)}{L^3} + \frac{\hat\Omega_{4}(E_0)}{L^4} + \frac{\hat\Omega_{5}(E_0)}{L^5} + \frac{\hat\Omega_{6}(E_0)}{L^6} + O(L^{-7})\, , \nonumber\\[2mm]
			\hat\Omega'(E_0) &=& \frac{\hat\Omega'_{2}(E_0)}{L^2} + \frac{\hat\Omega'_{3}(E_0)}{L^3} + O(L^{-4})\, , \nonumber\\[2mm]
			\hat\Omega''(E_0) &=& \hat\Omega''_0(E_0) +O(L^{-1})\, .
	\en
        Since $E-E_0=O(L^{-3})$, the terms, displayed above, are sufficient to calculate the energy
        shift up to and including order $L^{-6}$. Using iterations, one straightforwardly obtains
        \eq
        E-E_0&=&\frac{E_3}{L^3}+\frac{E_4}{L^4}+\frac{E_5}{L^5}+\frac{E_6}{L^6}+O(L^{-7})\, ,
\nonumber\\[2mm]
E_3&=&\hat\Omega_3(E_0)\, ,
\nonumber\\[2mm]
E_4&=&\hat\Omega_4(E_0)\, ,
\nonumber\\[2mm]
E_5&=&\hat\Omega_5(E_0)+\hat\Omega_2'(E_0)\hat\Omega_3(E_0)\, ,
\nonumber\\[2mm]
E_6&=&\hat\Omega_6(E_0)+\hat\Omega_3'(E_0)\hat\Omega_3(E_0)+\hat\Omega_2'(E_0)\hat\Omega_4(E_0)+\frac{1}{2}\,\hat\Omega_0''(E_0)\hat\Omega_3(E_0)^2\, .
        \en
 The matrix $\Omega(E)$ can be expanded similarly.
        The coefficients $\Omega_i(E_0),\Omega_i'(E_0),\Omega_i''(E_0)$ can be calculated by iterating
        Eq.~(\ref{eq:LS2}). Defining a $3\times 3$ matrix $V_{pq}$, whose
        elements are given by $\langle p|{\bf V}_{\alpha\beta}|q\rangle$, and attaching
        superscripts ``$c$,'' ``$d$,'' ``$r$,'' ``$\eta$,'' to the individual contributions,
        one gets a compact expression:
	\eq\label{eq:Omega}
		\frac{\Omega_{3}(E_0)}{L^3} &=& V^c_{00}, \nonumber\\[2mm]
		\frac{\Omega_{4}(E_0)}{L^4} &=& -\sum_{p \neq 0} \frac{V^c_{0p}V^c_{p0}}{E_p}, \nonumber\\[2mm]
		\frac{\Omega_{5}(E_0)}{L^5} &=& \sum_{p,k \neq 0} \frac{V^c_{0p}V^c_{pk}V^c_{k0}}{E_p E_k}, \nonumber\\[2mm] 
		\frac{\Omega_{6}(E_0)}{L^6} &=& -\sum_{p,k,q \neq 0} \frac{V^c_{0p}V^c_{pq}V^c_{qk}V^c_{k0}}{E_p E_q E_k} -\sum_{p \neq 0} \frac{V^c_{0p}V^d_{p0}+V^d_{0p}V^c_{p0}}{E_p} + V^\eta_{00} + \sum_{p,k \neq 0} \frac{V^c_{0p}V^r_{pk}V^c_{k0}}{E_p E_k}, \\[2mm]
		\phantom{25}  \nonumber\\[2mm]
                \label{eq:Omega1}
		\frac{\Omega'_{2}(E_0)}{L^2} &=& -\sum_{p \neq 0} \frac{V^c_{0p}V^c_{p0}}{E_p^2}, \nonumber\\[2mm]
		\frac{\Omega'_{3}(E_0)}{L^3} &=& 2\sum_{p,k \neq 0} \frac{V^c_{0p}V^c_{pk}V^c_{k0}}{E_p^2 E_k} ,\\[2mm]
		\phantom{25} \nonumber\\[2mm]
                \label{eq:Omega2}
		\Omega''_{0}(E_0) &=& 2\sum_{p \neq 0} \frac{V^c_{0p}V^c_{p0}}{E_p^3}.
	\en
        Taking the component $\alpha=\beta=1$ from the above expressions, we finally arrive at the expressions for
        $\hat\Omega_i(E_0)$, $\hat\Omega'_i(E_0)$, $\hat\Omega''_i(E_0)$, we are looking for.

        \medskip

        Putting everything together, we obtain the final expression for the energy shift:
  	\eq\label{eq:final}
E_3&=&\frac{4 \pi}{3 M} \left(5a_0 + 4a_2\right), \nonumber\\[2mm]
E_4&=&-\frac{4}{3 M} \left(5a_0^2 + 4a_2^2\right) I, \nonumber\\[2mm]
E_5&=& \frac{4}{3 M \pi} \biggl[ \left(5a_0^3 + 4a_2^3\right) I^2
  - \frac{1}{9} \left(55a_0^3 -120 a_0^2 a_2 -60 a_0 a_2^2 + 44 a_2^3\right) J  \biggr]\, , \nonumber\\[2mm]
E_6&=&-\frac{4}{3 M \pi^2} \Big[ \left(5a_0^4 +4a_2^4\right) I^3 - \frac{1}{9} \left(185a_0^4 -160 a_0^3 a_2 - 80 a_0 a_2^3 + 136 a_2^4\right) IJ  \nonumber\\[2mm]
			&-& \frac{1}{27} \left(5a_0 + 4a_2\right) \left(115a_0^3 + 420 a_0^2 a_2 + 300 a_0 a_2^2 + 56 a_2^3\right) K \nonumber\\[2mm]
			&+& \frac{8}{27}  \left(10a_0^4 + 140 a_0^3 a_2 + 285 a_0^2 a_2^2 +  50 a_0 a_2^3 + a_2^4\right) Q^r \nonumber\\[2mm]
			&+& \frac{8}{9}  \left(5a_0^4 + 20 a_0^3 a_2 + 45 a_0^2 a_2^2 +  10 a_0 a_2^3 + a_2^4\right) R^r \Big] \nonumber\\[2mm]
			&+&\frac{8\pi^2}{3M} \left(5a_0^3 r_0 + 4a_2^3 r_2\right) - \frac{4\pi^2}{3M^3} \left(5a_0^2 + 4 a_2^2\right) \nonumber\\[2mm]
			&+& \frac{64 \pi}{243 M}\Big[ 9\sqrt{3} \left(5a_0^4 + 20 a_0^3 a_2 + 45 a_0^2 a_2^2 +  10 a_0 a_2^3 + a_2^4\right) \nonumber\\[2mm]
			&-& 2\pi \left(10a_0^4 + 140 a_0^3 a_2 + 285 a_0^2 a_2^2 +  50 a_0 a_2^3 + a_2^4\right) \Big]\ln(\mu L)  \nonumber\\[2mm]
			&-&6 \eta_1^r - 10 \eta_2^r \, .
	\en
Finally, we give a list of all momentum sums that appear in the above equation. These
are the same quantities that enter the expression in case of the maximal isospin.
The quantities $I,J,K$ are finite in the dimensional regularization and
are given by
\eq
I&=&\sum_{{\bf n}\neq 0}\frac{1}{{\bf n}^2}=-8.91363291781\cdots\, , 
\nonumber\\[2mm]
J&=&\sum_{{\bf n}\neq 0}\frac{1}{{\bf n}^4}=16.532315959\cdots\, ,
\nonumber\\[2mm]
K&=&\sum_{{\bf n}\neq 0}\frac{1}{{\bf n}^6}=8.401923974433\cdots\, .
\en
The quantities $Q^r$ and $R^r$ represent the finite part of the double sums over momenta in
the minimal subtraction renormalization scheme. Introducing a formal notation
of a sum in $d$ dimensions,\footnote{The sums in $d$ dimensions can be interpreted as follows. One first performs the Poisson transform and singles out the infinite-volume part, which corresponds to an integral instead of a sum. The dimensional regularization and minimal subtraction is applied then to this integral. The remainder is ultraviolet-convergent and the limit $d\to 3$ can be performed there without a problem.} we get
\eq
\frac{1}{L^{2d}}\sum_{{\bf p},{\bf q}\neq 0}\frac{1}{{\bf p}^2{\bf q}^2({\bf p}^2+{\bf q}^2+({\bf p}+{\bf q})^2)}
=\mu^{2(d-3)}\biggl\{\frac{1}{48\pi^2}\,\biggl(\ln(\mu L)-\frac{1}{2(d-3)}
\biggr)
+\frac{1}{(2\pi)^6}\,Q^r\biggr\}\, ,
\nonumber\\[2mm]
\frac{1}{L^{2d}}\,\sum_{{\bf p}\neq 0}\frac{1}{{\bf p}^4}\,
\sum_{\bf q}\frac{1}{({\bf p}^2+{\bf q}^2+({\bf p}+{\bf q})^2)}
=\mu^{2(d-3)}\biggl\{-\frac{\sqrt{3}}{32\pi^3}\,\biggl(\ln(\mu L)-\frac{1}{2(d-3)}
\biggr)
+\frac{1}{(2\pi)^6}\,R^r\biggr\}\, .
\en
The finite parts are given by
\eq
Q^r&=&-102.1556055\cdots\, ,
\nonumber\\[2mm]
R^r&=&19.186903\cdots\, .
\en
Note that, in the above equations, ${\bf p}=2\pi{\bf n}/L$, ${\bf q}=2\pi{\bf m}/L$,
and the summation is carried out over ${\bf n},{\bf m}\in\mathbb{Z}^3$.

\medskip

Equation~(\ref{eq:final}) represents our final result. It gives the expression for the
energy shift of the ground state with $I=1$ up to and including $O(L^{-6})$.
Together with the known expression for the
case $I=3$, it provides a full framework for analyzing the lattice data on ground states
in the three-pion system. Two independent nonderivative couplings $\eta_{1,2}$ can be
separated and extracted from data in a result of this analysis.

\section{Conclusions}
\label{sec:concl}

\begin{itemize}

\item[i)]
  Using the nonrelativistic effective Lagrangian approach, we have evaluated the finite-volume
  shift of the ground state of three pions in the state with total isospin $I=1$ up to and including $O(L^{-6})$. Since three pions
  with zero momenta can have only $I=1$ or $I=3$, the expression, which is derived
  in the present paper, provides the last remaining missing piece in the description of the three-pion
  ground state, because the expression in the case of $I=3$ has been known in the
  literature. With this result at hand, one can also study the three-pion decays into
all isospin channels.  

\item[ii)]
  An immediate application of the obtained expression could be the extraction of two
  independent three-body couplings $\eta_{1,2}$ from the lattice data. Moreover, the
  analysis of the three-particle states might put constraints on the parameters in the
  two-particle sector -- the scattering lengths $a_I$ and the effective radii $r_I$. According
  to our past experience in the scalar $\varphi^4$-theory~\cite{Romero-Lopez:2018rcb,Romero-Lopez:2020rdq},
  these constraints are most effectively implemented when one performs a
  simultaneous fit of
  all datasets in the two- and three-particle sectors. The input from the three-particle sector might
  be interesting, because the two-particle scattering parameters (especially those corresponding to the total isospin $I=0$) are not very well determined on the lattice at present
  (for the recent developments in the two-particle sector, we refer the reader, e.g.,
  to~\cite{Yagi:2011jn,Fu:2013ffa,Helmes:2015gla,Liu:2016cba}).

\item[iii)]
  Albeit we are working in the nonrelativistic effective theory, our results contain a full
  set of relativistic corrections up to and including $O(L^{-6})$. In general, such corrections
  can be systematically included in the nonrelativistic approach up to a desired accuracy.

\item[iv)]  
  The above work is a rather straightforward generalization of the result obtained for
  the maximal isospin. One has, however, to deal with subtle issues, for example, with the degeneracy of levels, which are absent in the former case. The level splitting does not occur
  for the ground state, but one will have to face it, deriving expressions for the
  shifts of the excited states. A framework for the degenerate states can be worked
  out along the path
  which is pretty similar to the one considered in the present paper.
  The pertinent explicit formulae are, however,
  not  needed for the study of the ground state, and are therefore not displayed.

\end{itemize}

\section*{Acknowledgments}

The authors would like to thank U.-G.~Mei{\ss}ner, F.~Romero-L\'opez and C. Urbach
for interesting discussions.
The work of F.~M. and  A.~R.  was  supported in part by the DFG (CRC 110 ``Symmetries 
and the Emergence of Structure in QCD'', Grant no. TRR110). A.~R., in addition,
thanks Volkswagenstiftung 
(Grant no. 93562) and the Chinese Academy of Sciences (CAS) President's
International Fellowship Initiative (PIFI) (Grant no. 2021VMB0007) for the partial
financial support.


\begin{thebibliography}{99}
  
\bibitem{Beane:2007es}
S.~R.~Beane, W.~Detmold, T.~C.~Luu, K.~Orginos, M.~J.~Savage and A.~Torok,
Phys. Rev. Lett. \textbf{100} (2008) 082004
[arXiv:0710.1827 [hep-lat]].

\bibitem{Horz:2019rrn}
B.~H\"orz and A.~Hanlon,
Phys. Rev. Lett. \textbf{123} (2019) 142002
[arXiv:1905.04277 [hep-lat]].

\bibitem{Culver:2019vvu}
C.~Culver, M.~Mai, R.~Brett, A.~Alexandru and M.~D\"oring,
Phys. Rev. D \textbf{101} (2020) 114507
[arXiv:1911.09047 [hep-lat]].

\bibitem{Fischer:2020jzp}
M.~Fischer, B.~Kostrzewa, L.~Liu, F.~Romero-L\'opez, M.~Ueding and C.~Urbach,
[arXiv:2008.03035 [hep-lat]].

\bibitem{Blanton:2019vdk}
T.~D.~Blanton, F.~Romero-L\'opez and S.~R.~Sharpe,
Phys. Rev. Lett. \textbf{124} (2020) 032001
[arXiv:1909.02973 [hep-lat]].

\bibitem{Hansen:2020otl}
M.~T.~Hansen, R.~A.~Brice\~no, R.~G.~Edwards, C.~E.~Thomas and D.~J.~Wilson,
Phys. Rev. Lett. \textbf{126} (2021) 012001
[arXiv:2009.04931 [hep-lat]].

\bibitem{Alexandru:2020xqf}
A.~Alexandru, R.~Brett, C.~Culver, M.~D\"oring, D.~Guo, F.~X.~Lee and M.~Mai,
Phys. Rev. D \textbf{102} (2020) 114523
[arXiv:2009.12358 [hep-lat]].


\bibitem{Romero-Lopez:2018rcb}
F.~Romero-L\'opez, A.~Rusetsky and C.~Urbach,
Eur. Phys. J. C \textbf{78} (2018) 846
[arXiv:1806.02367 [hep-lat]].

\bibitem{Romero-Lopez:2020rdq}
F.~Romero-L\'opez, A.~Rusetsky, N.~Schlage and C.~Urbach,
JHEP \textbf{60} (2021) 60
[arXiv:2010.11715 [hep-lat]].


\bibitem{Hansen:2014eka}
  M.~T.~Hansen and S.~R.~Sharpe,
  Phys.\ Rev.\ D {\bf 90} (2014) 116003 [arXiv:1408.5933 [hep-lat]].

\bibitem{Hansen:2015zga}
  M.~T.~Hansen and S.~R.~Sharpe,
  Phys.\ Rev.\ D {\bf 92} (2015) 114509 [arXiv:1504.04248 [hep-lat]].




\bibitem{Polejaeva:2012ut}
  K.~Polejaeva and A.~Rusetsky,
  Eur.\ Phys.\ J.\ A {\bf 48} (2012) 67 [arXiv:1203.1241 [hep-lat]].


\bibitem{Hammer:2017uqm}
H.~W.~Hammer, J.~Y.~Pang and A.~Rusetsky,
JHEP \textbf{09} (2017) 109
[arXiv:1706.07700 [hep-lat]].


\bibitem{Hammer:2017kms}
H.~W.~Hammer, J.~Y.~Pang and A.~Rusetsky,
JHEP \textbf{10} (2017) 115
[arXiv:1707.02176 [hep-lat]].



\bibitem{Mai:2017bge}
  M.~Mai and M.~D\"oring,
  Eur.\ Phys.\ J.\ A {\bf 53} (2017)  240
  [arXiv:1709.08222 [hep-lat]].

\bibitem{Hansen:2015zta}
  M.~T.~Hansen and S.~R.~Sharpe,
  Phys.\ Rev.\ D {\bf 93} (2016) 014506 [arXiv:1509.07929 [hep-lat]].

\bibitem{Hansen:2016fzj}
  M.~T.~Hansen and S.~R.~Sharpe,
  Phys.\ Rev.\ D {\bf 93} (2016) 096006 [arXiv:1602.00324 [hep-lat]].


\bibitem{Briceno:2017tce}
  R.~A.~Brice\~no, M.~T.~Hansen and S.~R.~Sharpe,
  Phys.\ Rev.\ D {\bf 95} (2017)  074510
  [arXiv:1701.07465 [hep-lat]].



\bibitem{Sharpe:2017jej}
  S.~R.~Sharpe,
  Phys.\ Rev.\ D {\bf 96} (2017)  054515
  [arXiv:1707.04279 [hep-lat]].


\bibitem{Briceno:2018mlh}
R.~A.~Brice\~no, M.~T.~Hansen and S.~R.~Sharpe,
Phys. Rev. D \textbf{98} (2018) 014506
[arXiv:1803.04169 [hep-lat]].

\bibitem{Briceno:2018aml}
R.~A.~Brice\~no, M.~T.~Hansen and S.~R.~Sharpe,
Phys. Rev. D \textbf{99} (2019) 014516
[arXiv:1810.01429 [hep-lat]].

\bibitem{Blanton:2019igq}
T.~D.~Blanton, F.~Romero-L\'opez and S.~R.~Sharpe,
JHEP \textbf{03} (2019) 106
[arXiv:1901.07095 [hep-lat]].



\bibitem{Briceno:2019muc}
R.~A.~Brice\~no, M.~T.~Hansen, S.~R.~Sharpe and A.~P.~Szczepaniak,
Phys. Rev. D \textbf{100} (2019) 054508
[arXiv:1905.11188 [hep-lat]].

\bibitem{Romero-Lopez:2019qrt}
F.~Romero-L\'opez, S.~R.~Sharpe, T.~D.~Blanton, R.~A.~Brice\~no and M.~T.~Hansen,
JHEP \textbf{10} (2019) 007
[arXiv:1908.02411 [hep-lat]].




\bibitem{Hansen:2020zhy}
M.~T.~Hansen, F.~Romero-L\'opez and S.~R.~Sharpe,
JHEP \textbf{07} (2020) 047
[arXiv:2003.10974 [hep-lat]].
  

\bibitem{Meissner:2014dea}
  U.-G.~Mei{\ss}ner, G.-Rios and A.~Rusetsky,
  Phys.\ Rev.\ Lett.\  {\bf 114} (2015) 091602,
   Erratum: [Phys.\ Rev.\ Lett.\  {\bf 117} (2016) 069902]
[arXiv:1412.4969 [hep-lat]].

\bibitem{Blanton:2020jnm}
T.~D.~Blanton and S.~R.~Sharpe,
Phys. Rev. D \textbf{102} (2020) 054515
[arXiv:2007.16190 [hep-lat]].

\bibitem{Blanton:2020gha}
T.~D.~Blanton and S.~R.~Sharpe,
Phys. Rev. D \textbf{102} (2020)  054520
[arXiv:2007.16188 [hep-lat]].

\bibitem{Doring:2018xxx}
M.~D\"oring, H.~W.~Hammer, M.~Mai, J.~Y.~Pang, A.~Rusetsky and J.~Wu,
Phys. Rev. D \textbf{97} (2018) 114508
[arXiv:1802.03362 [hep-lat]].
  

\bibitem{Pang:2019dfe}
J.~Y.~Pang, J.~J.~Wu, H.~W.~Hammer, U.-G.~Mei\ss{}ner and A.~Rusetsky,
Phys. Rev. D \textbf{99} (2019)  074513
[arXiv:1902.01111 [hep-lat]].

\bibitem{Pang:2020pkl}
J.~Y.~Pang, J.~J.~Wu and L.~S.~Geng,
Phys. Rev. D \textbf{102} (2020) 114515
[arXiv:2008.13014 [hep-lat]].


\bibitem{Jackura:2019bmu}
A.~W.~Jackura, S.~M.~Dawid, C.~Fern\'andez-Ram\'\i{}rez, V.~Mathieu, M.~Mikhasenko, A.~Pilloni, S.~R.~Sharpe and A.~P.~Szczepaniak,
Phys. Rev. D \textbf{100} (2019)  034508
[arXiv:1905.12007 [hep-ph]].








\bibitem{Meng:2017jgx}
Y.~Meng, C.~Liu, U.-G.~Mei\ss{}ner and A.~Rusetsky,
Phys. Rev. D \textbf{98} (2018) 014508
[arXiv:1712.08464 [hep-lat]].




  \bibitem{Mai:2018djl}
M.~Mai and M.~D\"oring,
Phys. Rev. Lett. \textbf{122} (2019) 062503
[arXiv:1807.04746 [hep-lat]].



\bibitem{Mai:2019fba}
M.~Mai, M.~D\"oring, C.~Culver and A.~Alexandru,
Phys. Rev. D \textbf{101} (2020)  054510
[arXiv:1909.05749 [hep-lat]].

\bibitem{Konig:2017krd}
S.~K\"onig and D.~Lee,
Phys. Lett. B \textbf{779} (2018) 9
[arXiv:1701.00279 [hep-lat]].

\bibitem{Klos:2018sen}
P.~Klos, S.~K\"onig, H.~W.~Hammer, J.~E.~Lynn and A.~Schwenk,
Phys. Rev. C \textbf{98} (2018)  034004
[arXiv:1805.02029 [nucl-th]].

\bibitem{Konig:2020lzo}
S.~K\"onig,
Few Body Syst. \textbf{61} (2020)  20
[arXiv:2005.01478 [hep-lat]].

\bibitem{Blanton:2020gmf}
T.~D.~Blanton and S.~R.~Sharpe,
[arXiv:2011.05520 [hep-lat]].



\bibitem{Kreuzer:2010ti}
  S.~Kreuzer and H.-W.~Hammer,
  Phys.\ Lett.\ B {\bf 694} (2011) 424 [arXiv:1008.4499 [hep-lat]].

\bibitem{Kreuzer:2009jp}
 S.~Kreuzer and H.-W.~Hammer,
  Eur.\ Phys.\ J.\ A {\bf 43} (2010) 229 [arXiv:0910.2191 [nucl-th]].

\bibitem{Kreuzer:2008bi}
  S.~Kreuzer and H.-W.~Hammer,
  Phys.\ Lett.\ B {\bf 673} (2009) 260 [arXiv:0811.0159 [nucl-th]].

\bibitem{Kreuzer:2012sr}
  S.~Kreuzer and H.-W.~Grie{\ss}hammer,
  Eur.\ Phys.\ J.\ A {\bf 48} (2012) 93 [arXiv:1205.0277 [nucl-th]].



\bibitem{Hansen:2019nir}
M.~T.~Hansen and S.~R.~Sharpe,
Ann. Rev. Nucl. Part. Sci. \textbf{69} (2019) 65
[arXiv:1901.00483 [hep-lat]].


\bibitem{Lee:1957zzb}
T.~D.~Lee, K.~Huang and C.~N.~Yang,
Phys. Rev. \textbf{106} (1957) 1135.

  
  
\bibitem{Huang:1957im}
  K.~Huang and C.~N.~Yang,
  Phys.\ Rev.\  {\bf 105} (1957) 767.


\bibitem{Wu:1959zz}
T.~T.~Wu,
Phys. Rev. \textbf{115} (1959) 1390.


\bibitem{Tan:2007bg}
S.~Tan,
Phys. Rev. A \textbf{78} (2008) 013636
[arXiv:0709.2530 [cond-mat.stat-mech]].

\bibitem{Beane:2007qr}
S.~R.~Beane, W.~Detmold and M.~J.~Savage,
Phys. Rev. D \textbf{76} (2007) 074507
[arXiv:0707.1670 [hep-lat]].



\bibitem{Detmold:2008gh}
  W.~Detmold and M.~J.~Savage,
  Phys.\ Rev.\ D {\bf 77} (2008) 057502
  [arXiv:0801.0763 [hep-lat]].


\bibitem{Beane:2020ycc}
S.~R.~Beane, W.~Detmold, R.~Horsley, M.~Illa, M.~Jafry, D.~J.~Murphy, Y.~Nakamura, H.~Perlt, P.~E.~L.~Rakow and G.~Schierholz, \textit{et al.}
[arXiv:2003.12130 [hep-lat]].




\bibitem{Yagi:2011jn}
T.~Yagi, S.~Hashimoto, O.~Morimatsu and M.~Ohtani,
[arXiv:1108.2970 [hep-lat]].

\bibitem{Fu:2013ffa}
Z.~Fu,
Phys. Rev. D \textbf{87} (2013)  074501
[arXiv:1303.0517 [hep-lat]].

\bibitem{Helmes:2015gla}
C.~Helmes \textit{et al.} [ETM],
JHEP \textbf{09} (2015) 109
[arXiv:1506.00408 [hep-lat]].

\bibitem{Liu:2016cba}
L.~Liu, S.~Bacchio, P.~Dimopoulos, J.~Finkenrath, R.~Frezzotti, C.~Helmes, C.~Jost, B.~Knippschild, B.~Kostrzewa and H.~Liu, \textit{et al.}
Phys. Rev. D \textbf{96} (2017)  054516
[arXiv:1612.02061 [hep-lat]].


  
\end{thebibliography}
\end{document}